\documentclass[aps,prb,twocolumn,superscriptaddress]{revtex4-2}
\usepackage{graphicx}
\usepackage{latexsym}
\usepackage{amssymb}
\usepackage{amsmath}
\usepackage{amsfonts}
\usepackage[vcentermath]{youngtab}
\usepackage{upgreek}
\usepackage{bm,dsfont}
\usepackage{multirow}
\usepackage{enumitem}
\usepackage[usenames, dvipsnames]{color}
\usepackage[svgnames]{xcolor}
\usepackage{hyperref}
\hypersetup{
colorlinks = true,
linkcolor = [rgb]{0.70,0.13,0.13},
citecolor = [rgb]{0.13,0.55,0.13},
urlcolor = [rgb]{0.25, 0.41, 0.88}}
\usepackage{makecell}

\newcommand{\ket}[1]{|#1 \rangle}

\newcommand{\dd}{\mathrm{d}}

\newcommand{\ii}{\mathrm{i}}
\newcommand{\e}{\mathrm{e}}

\newcommand{\LU}{\mathrm{LU}}
\newcommand{\U}{\mathrm{U}}

\newcommand{\dsR}{\mathbb{R}}

\newcommand{\dsZ}{\mathbb{Z}}

\newcommand{\scL}{\mathcal{L}}

\newcommand{\scV}{\mathcal{V}}

\newcommand{\sgn}{\operatorname{sgn}}

\newcommand{\vect}[1]{{\bm{#1}}}

\newcommand{\mat}[1]{\left[\begin{matrix}#1\end{matrix}\right]}
\newcommand{\smat}[1]{\left[\begin{smallmatrix}#1\end{smallmatrix}\right]}
\newcommand{\eq}[1]{\begin{equation}#1\end{equation}}
\newcommand{\eqs}[1]{\begin{equation}\begin{split}#1\end{split}\end{equation}}
\newcommand{\eqnref}[1]{Eq.\,\eqref{#1}}
\newcommand{\figref}[1]{Fig.\,\ref{#1}}
\newcommand{\tabref}[1]{Tab.\,\ref{#1}}
\newcommand{\secref}[1]{Sec.\,\ref{#1}}
\newcommand{\appref}[1]{Appendix\,\ref{#1}}
\newcommand{\refcite}[1]{Ref.\,\onlinecite{#1}}

\newcommand{\bea}{\begin{eqnarray}}
\newcommand{\eea}{\end{eqnarray}}
\def\be{\begin{equation}}
\def\ee{\end{equation}}
\newcommand{\beq}{\begin{equation}}
\newcommand{\eeq}{\end{equation}}
\newcommand{\beqn}{\begin{eqnarray}}
\newcommand{\eeqn}{\end{eqnarray}}

\newcommand{\hda}{H_\text{int, AA}}
\newcommand{\hdb}{H_\text{int, AB}}
\newcommand{\hbb}{H_\text{int, BB}}
\newcommand{\hcf}{H_\text{int, CF}}

\newcommand{\abs}[1]{|#1|}

\usepackage{ulem}
\begin{document}

\title{Fermi Surface Symmetric Mass Generation}

\author{Da-Chuan Lu}
\affiliation{Department of Physics, University of California, San Diego, CA 92093, USA}
\author{Meng Zeng}
\affiliation{Department of Physics, University of California, San Diego, CA 92093, USA}
\author{Juven Wang}
\affiliation{Center of Mathematical Sciences and Applications, Harvard University, Cambridge, {MA 02138}, USA}
\author{Yi-Zhuang You}
\affiliation{Department of Physics, University of California, San Diego, CA 92093, USA}


\begin{abstract}
Symmetric mass generation is a novel mechanism to give gapless fermions a mass gap by non-perturbative interactions without generating any fermion bilinear condensation. The previous studies of symmetric mass generation have been limited to Dirac/Weyl/Majorana fermions with zero Fermi volume in the free fermion limit. In this work, we generalize the concept of symmetric mass generation to Fermi liquid with a finite Fermi volume and discuss how to gap out the Fermi surfaces by interactions without breaking the U(1) loop group symmetry or developing topological orders. We provide examples of Fermi surface symmetric mass generation in both (1+1)D and (2+1)D Fermi liquid systems when several Fermi surfaces together cancel the Fermi surface anomaly. However, the U(1) loop group symmetry in these cases is still restrictive enough to rule out all possible fermion bilinear gapping terms, such that a non-perturbative interaction mechanism is the only way to gap out the Fermi surfaces. This symmetric Fermi surface reconstruction is in contrast to the conventional symmetry-breaking mechanism to gap the Fermi surfaces. As a side product, our model provides a pristine 1D lattice regularization for the (1+1)D U(1) symmetric chiral fermion model (e.g., the 3-4-5-0 model) by utilizing a lattice translation symmetry as an emergent U(1) symmetry at low energy. This opens up the opportunity for efficient numerical simulations of chiral fermions in their own dimensions \emph{without} introducing mirror fermions under the domain wall fermion construction.
\end{abstract}
\maketitle

\section{Introduction}

Fermi liquids are gapless quantum many-body systems of fermions that possess Fermi surfaces and well-defined quasi-particle excitations at low energy. They are the models for the most commonly-seen metallic materials in nature. They are probably also one of the most studied quantum phases of matter in condensed matter physics since Landau 
\cite{landau2013statistical, lifshitz2013statistical}. However, there are still many novel aspects of Fermi liquids that might not have been well recognized. This article explores one such aspect: the phenomenon of \emph{symmetric mass generation} (SMG, see a recent overview \cite{Wang2204.14271} and references therein) in Fermi liquids.

One intriguing property of the Fermi liquid is the surprising stability of the Fermi surface under generic local interactions of fermions.  
Although the system is gapless with vastly degenerated ground states, local interactions often do not immediately lift the ground state degeneracy and destabilize the Fermi liquid towards gapped phases. 
Early understandings of this property came from the perturbative renormalization group (RG) analysis, as the Fermi liquid theory can emerge as a stable RG fix-point of interacting fermion systems \cite{Shankar1991, Shankarcond-mat/9307009, Hewsoncond-mat/9410013, Chitovcond-mat/9505140, Dupuiscond-mat/9511120, Chitovcond-mat/9705037, Rademaker1507.07276}.

Recently, a modern understanding arose under the name of \emph{Fermi surface anomaly} \cite{Else2007.07896, Else2010.10523}, which states that the stability of the Fermi surface can be viewed as protected by the quantum anomaly of an emergent $\LU(1)$  loop group symmetry at low energy, extending and unifying many related discussions \cite{Oshikawacond-mat/9610168, Oshikawacond-mat/9911137, Oshikawacond-mat/0002392, Misguichcond-mat/0112360, Paramekanticond-mat/0406619, Haldanecond-mat/0505529, Hastingscond-mat/0411094, Watanabe1505.04193, Cheng1511.02263, Lu1705.09298, Cho1705.03892, Jian1705.00012, Metlitski1707.07686, Bultinck1808.00324, Song1909.08637, Yao1906.11662} about Luttinger's theorem \cite{LuttingerRP1960} and Lieb-Schultz-Mattis theorem \cite{LSM1961}. Loosely speaking, the $\LU(1)$ symmetry corresponds to the fermion number $n_\vect{k}$ conservation at each momentum point $\vect{k}$ on the Fermi surface (FS), which is preserved by the Landau Fermi liquid Hamiltonian $H_\text{FL}=\sum_{\vect{k}\in\text{FS}}\epsilon_\vect{k}n_\vect{k}+\sum_{\vect{k},\vect{k'}\in\text{FS}}f_{\vect{k}\vect{k}'}n_\vect{k}n_{\vect{k}'}+\cdots$. In the presence of the Fermi surface anomaly, the Fermi liquid can only be gapped by either (i) spontaneously breaking the $\LU(1)$ symmetry or (ii) spontaneously developing anomalous topological orders (or other non-Fermi liquid exotic states) that saturate the Fermi surface anomaly. The anomaly matching is a kinematic constraint, which is non-perturbative and more robust than the perturbative RG analysis of the Fermi liquid low-energy dynamics.

Over the past decade, the quantum anomaly \cite{Ryu1010.0936, Wen1303.1803, Kapustin1404.3230, Wang1405.7689} has been realized as an important theoretical tool in analyzing the protected gapless boundary states of interacting topological insulators/superconductors, which belong to symmetry-protected topological (SPT) phases 
in a grand scope (see overviews
\cite{Senthil1405.4015, Witten1508.04715, Wen1610.03911} 
and references therein). An interesting phenomenon, known as symmetric mass generation (SMG) 
\cite{Fidkowski0904.2197, Fidkowski1008.4138, Ryu1202.4484, Qi1202.3983, Yao1202.5805, Gu1304.4569,
Wang1307.7480, Slagle1409.7401, Ayyar1410.6474, Catterall1510.04153, Ayyar1511.09071}, was discovered in the study of interacting fermionic SPT states. 
It was realized that certain SPT states might look non-trivial at the free-fermion (non-interacting) level but can be smoothly deformed into a trivial gapped phase with a unique ground state by fermion interactions. This implies some integer $\mathbb{Z}$ classification of non-interacting SPT states can be reduced to a finite abelian elementary order-$n$ group $\mathbb{Z}_n$ classification for some interacting SPT states, first emphasized by Fidkowski-Kitaev \cite{Fidkowski0904.2197, Fidkowski1008.4138}. Correspondingly, their gapless boundary states can be  gapped out by (and only by) interaction without breaking the symmetry or developing the topological order (breaking emergent higher-form symmetry). This provides a novel mechanism to generate a mass for zero-density relativistic gapless fermions 
(e.g., Dirac/Weyl/Majorana fermions occupying only Fermi points with zero Fermi volumes at the Fermi level, colloquially known as Dirac/Weyl/Majorana cones) without symmetry breaking. This mechanism is called SMG, which is distinct from the conventional Higgs mechanism that relies on symmetry breaking for fermion mass generation. 

However, as far as we are aware of the existing literature,
the SMG mechanism has not yet been extended to 
fermion systems at a finite filling (with a finite density). The Fermi liquid is one most notable examples of such, which possesses a Fermi surface enclosing a finite Fermi volume. It is natural to ask: can SMG happen on the Fermi surface as well, gapping out the Fermi surface by interaction without breaking the loop group symmetry of interest? As we will demonstrate in this article, the answer is yes. 

Given the spacetime-internal symmetry $G$ of a fermion system, the conditions \cite{Wang2204.14271} for SMG to happen are: (i) the system must be free from $G$-anomaly such that symmetric gapping (without topological order) becomes possible, and (ii) the symmetry $G$ must be restricted enough to rule out any symmetric fermion bilinear gapping term such that the gapping can only be achieved by interaction. These defining conditions of SMG can be applied to the Fermi liquid system by considering $G$ as the emergent loop group symmetry on the Fermi surface. Based on this understanding, we will investigate the Fermi surface SMG in the presence of the $\LU(1)$ symmetry. The general feature is that even though a single Fermi surface is anomalous, it is possible to cancel the Fermi surface anomaly among multiple Fermi surfaces (or Fermi surfaces with multiple fermion flavors), such that interactions can drive the transition from the Fermi liquid phase to a symmetric gapped phase. We shall name this phenomenon
as the ``Fermi surface Symmetric Mass Generation''.

The Fermi surface SMG provides us a different possibility to create a gap to all excitations on the Fermi surface without condensing any fermion bilinear order parameter, which makes it distinct from the superconducting gap (i.e., condensing Cooper pairs) or the density wave gap (i.e., condensing excitons) that are more familiar in condensed matter physics. Nevertheless, it does involve condensing some multi-fermion bound states that transform trivially under the symmetry action. One simplest example is the charge-4e superconductor \cite{KivelsonPRB1990, Kameicond-mat/0505468, Berg0810.1564, Radzihovsky0812.3945, Berg0904.1230, Moon1202.5389, Jiang1607.01770}, which condenses fermion quartets (four-fermion bound states) that preserves at least the $\dsZ_4$ subgroup of the charge $\U(1)$ symmetry. In this work, we will provide more carefully designed examples preserving the full $\U(1)$ symmetry (and other lattice symmetries), but the essential idea of condensing symmetric multi-fermion operators to generate a many-body excitation gap is the same. Therefore, the Fermi surface SMG is intrinsically a strong non-perturbative interaction effect of fermions. The interaction may look irrelevant at the free-fermion (or the Fermi liquid) fixed-point. However, strong enough interaction can still drive the gap-opening transition through non-perturbative effects.

The article will be organized as follows. In \secref{sec:1+1}, we will present a lattice model of Fermi surface SMG in (1+1)D, as the pristine lattice regularization of the 3-4-5-0 chiral fermion model, whose phase diagram can be reliably analyzed by RG approach. In \secref{sec:2+1}, we will extend the discussion of Fermi surface SMG to (2+1)D in a concrete lattice model, which can be exactly solved in both the weak and strong interaction limits. Through these examples, we establish the Fermi surface SMG as a general mechanism to gap out anomaly-free Fermi surfaces in different dimensions. We summarize our result and discuss its connection to future directions in \secref{sec:sum}.

\section{Fermi Surface SMG in (1+1)D}\label{sec:1+1}

\subsection{(1+1)D Fermi Liquid and Fermi Surface Anomaly}

\begin{figure}[htbp]
\begin{center}
\includegraphics[scale=0.65]{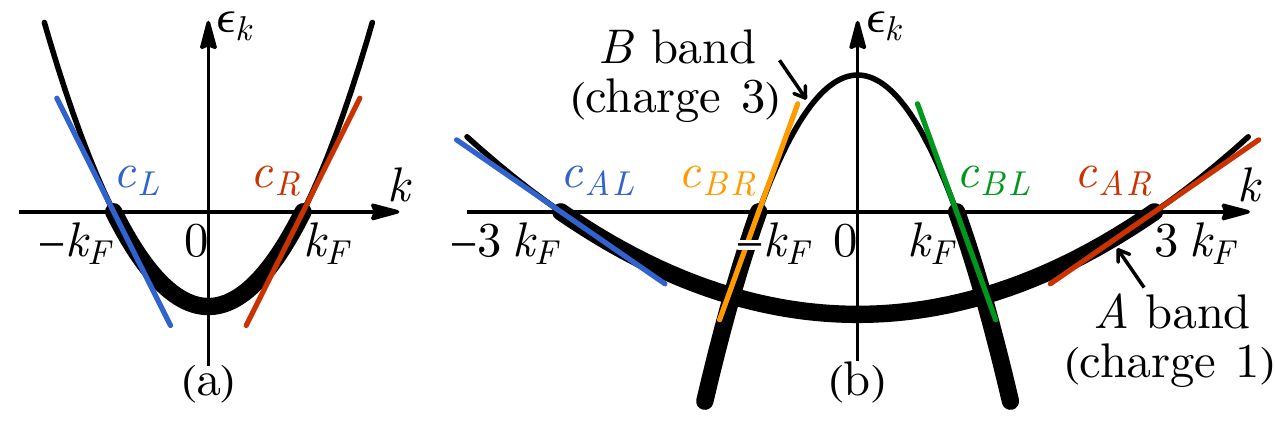}
\caption{(a) A typical single-band Fermi liquid with Fermi surface anomaly. (b) Two-band model of a Fermi liquid with the Fermi surface anomaly canceled. Chiral fermions with linearized dispersions around different Fermi points emerge at low energy.}
\label{fig: dispersion}
\end{center}
\end{figure}

In the free-fermion limit, the (1+1)D Fermi liquid can be realized as a system of fermions occupying a segment of single-particle momentum eigenstates in the 1D momentum space (or Brillouin zone), which can be described by a Hamiltonian $H=\sum_{k}c_k^\dagger \epsilon_k c_k$, where $c_k$ (or $c_k^\dagger$) is the fermion annihilation (or creation) operator of the single-particle mode at momentum $k$. For now, we only consider spinless fermions, such that the $c_{k}$ operator does not carry spin (or any other internal degrees of freedom). As an example, suppose the band structure is described by $\epsilon_{k}=(k^2-k_{F}^2)/(2m)$ for non-relativistic fermions with a finite chemical potential $\mu=k_{F}^2/(2m)$. The ground state of the Hamiltonian $H$ will have fermions occupying the momentum segment $k\in[-k_{F},k_{F}]$ set by the Fermi momentum $k_F$, as illustrated in \figref{fig: dispersion}(a).

The low-energy degrees of freedom in the (1+1)D Fermi liquids can be modeled by the chiral fermions near the 0D Fermi surfaces 
(namely, Fermi points) at $\pm k_F$, which are described by the following Lagrangian density
\eq{\label{eq: L FL}
\scL=c_L^\dagger(\ii\partial_t-v_F\ii\partial_x)c_L+c_R^\dagger(\ii\partial_t+v_F\ii\partial_x)c_R,}
where $v_F=k_F/m$ is the Fermi velocity. 
The operator $c_L$ (or $c_R$) annihilates the left (or right) moving fermion modes, defined as \eq{c_{R/L}(x)=\int_{-\Lambda}^{\Lambda}\dd\kappa\;c_{\pm k_F+\kappa} \e^{\ii(\pm k_F+\kappa)x}} 
around the Fermi points within a small momentum cutoff $\Lambda\ll k_F$. The low-energy effective theory $\scL$ in \eqnref{eq: L FL} has an emergent $\U(1)_L\times\U(1)_R$ symmetry, corresponding to the separate charge conservation of the left- and right-moving chiral fermions. Under the symmetry transformation with the periodic $\phi_L$ and $\phi_R$ in $[0, 2 \pi)$,
\eqs{
\U(1)_L&:c_L\to\e^{\ii\phi_L}c_L,c_R\to c_R;\\
\U(1)_R&:c_L\to c_L,c_R\to \e^{\ii\phi_R}c_R.}
They can be as well understood as a recombination of the vector $\U(1)_V$ and axial $\U(1)_A$ symmetries,
by rewriting $\phi_L=\phi-k_F \,\delta x$ and $\phi_R=\phi+k_F \,\delta x$,
\eqs{
\U(1)_V&:c_k\to \e^{\ii\phi}c_k\Rightarrow
\left\{\begin{array}{l}
c_L\to\e^{\ii\phi}c_L,\\
c_R\to \e^{\ii\phi}c_R;\end{array}\right.\\
\U(1)_A&:c_k\to \e^{\ii k \,\delta x}c_k \Rightarrow\left\{\begin{array}{l}
c_L\to \e^{-\ii k_F \,\delta x}c_L,\\
c_R\to \e^{+\ii k_F \,\delta x}c_R.\end{array}\right.}
More precisely, the combined symmetry group should be denoted as $\U(1)_V\times_{\dsZ_2^F}\U(1)_A\equiv \frac{\U(1)_V\times \U(1)_A}{\dsZ_2^F}$, because  the $\U(1)_V$ and $\U(1)_A$ symmetries share the fermion parity $\dsZ_2^F$ subgroup (under which $c_{L,R}\to - c_{L,R}$).
The physical meaning of the vector $\U(1)_V$ symmetry is the total $\U(1)$ charge conservation of the fermions, and the axial $\U(1)_A$ symmetry can be considered an effective representation of the \emph{translation symmetry} in the infrared (IR) limit (that translates all fermions by displacement $\delta x$ along the 1D system). Although translation symmetry is described by a \emph{non-compact} symmetry group $\dsZ$ at the lattice scale, its action on the low-energy chiral fermion fields $c_L,c_R$ behaves as a \emph{compact} $\U(1)_A$ emergent symmetry \cite{ Metlitski1707.07686, Darius-Shi2204.07585}.

The stability of the Fermi liquid is protected by the Fermi surface anomaly, which can be viewed as the mixed anomaly between the $\U(1)_V$ and $\U(1)_A$ symmetries. The anomaly index is given by \cite{LuttingerRP1960, LSM1961, Cho1705.03892}
\eq{1\times k_F-1\times(-k_F)=2k_F=2\pi\nu,}
which can be related to the fermion filling fraction $\nu$. The system is anomalous if the filling $\nu$ is not an integer. Without breaking the charge $\U(1)$ and translation symmetries, it is impossible to drive the Fermi liquid to a trivial gap phase due to the non-vanishing Fermi surface anomaly. This can be viewed as a consequence of the Lieb-Schultz-Mattis (LSM) theorem \cite{LSM1961}. The situation is also similar to the chiral fermion edge states on the (1+1)D boundary of a (2+1)D quantum Hall insulator.

\subsection{Two-Band Model and Anomaly Cancellation}

To generate a gap for these low-energy fermions in (1+1)D Fermi liquids, the Fermi surface anomaly must be canceled. Here we present a two-band toy model that achieves anomaly cancellation and enables gapping out the Fermi surface without breaking the charge $\U(1)$ and translation symmetries and without generating any Fermi bilinear condensation. It will provide a concrete example of SMG in (1+1)D Fermi liquids.

Consider a 1D lattice (a chain of sites) with two types of fermions $c_{iA}$ and $c_{iB}$ per site. The $A$-type fermion $c_{iA}$ carries charge $q_A$ under a global $\U(1)$ symmetry, and the $B$-type fermion $c_{iB}$ carries charge $q_B$ under the same $\U(1)$ symmetry. The Hamiltonian takes the general form of
\eqs{\label{eq: H two-band}
H=&-\sum_{ij}(t_{ij}^Ac_{iA}^\dagger c_{jA} + t_{ij}^Bc_{iB}^\dagger c_{jB}+\text{h.c.})\\
&-\sum_{i}(\mu_A c_{iA}^\dagger c_{iA}+\mu_B c_{iB}^\dagger c_{iB})+H_\text{int}}
with $H_\text{int}$ being some fermion interactions to be specified later in \eqnref{eq: H two-band interaction}. The specific details of the hopping coefficients $t_{ij}^A$ and $t_{ij}^B$ are not important to our discussion as long as they produce a band structure that looks like \figref{fig: dispersion}(b) in the Brillouin zone. The $A$-type fermion forms an electron-like band, and the $B$-type fermion forms a hole-like band. The two bands overlap in the energy spectrum. This will realize a two-band Fermi liquid in general. The Hamiltonian $H$ in \eqnref{eq: H two-band} has a $\U(1)\times(\dsZ\rtimes\dsZ_2)$ symmetry (parameterized by a periodic angle $\phi\in[0,2\pi)$ and an integer $n\in\dsZ$ as follows)
\eqs{\label{eq: latt symm}
\U(1)&:c_{iA}\to\e^{\ii q_A\phi}c_{iA},\quad c_{iB}\to\e^{\ii q_B\phi}c_{iB};\\
\dsZ&:c_{iA}\to c_{(i+n)A},\quad c_{iB}\to c_{(i+n)B};\\
\dsZ_2&:c_{iA}\to c_{(-i)A},\quad c_{iB}\to c_{(-i)B}.}
They correspond to the total charge conservation symmetry $\U(1)$, the lattice translation symmetry $\dsZ$, and the lattice reflection symmetry $\dsZ_2$. The question is whether we can gap the Fermi liquid without breaking all these symmetries in (1+1)D.

One significant obstruction towards gapping is the Fermi surface anomaly, which can also be interpreted as a mixed anomaly between the charge $\U(1)$ and (the IR correspondence of) the translation symmetry. To cancel the Fermi surface anomaly, we need to fine-tune the chemical potentials $\mu_A$ and $\mu_B$ such that the anomaly index vanishes
\eq{\label{eq: anomaly cancellation}
q_A \nu_A + q_B \nu_B=0\mod 1,}
where $\nu_A$ and $\nu_B$ are the filling fractions of the $A$ and $B$ bands (for the hole-like $B$ band, we may assign $\nu_B<0$ such that $|\nu_B|$ corresponds to the hole-filling). This is also known as the \emph{charge compensation} condition in semiconductor physics.

If the $A$-type and $B$-type fermions carry the same charge as $q_A=q_B=1$, the anomaly cancellation condition \eqnref{eq: anomaly cancellation} simply requires $\nu_A=-\nu_B$. In this case, the electron-like Fermi surface of the $A$-type fermion and the hole-like Fermi surface of the $B$-type fermion are perfectly nested (with zero nesting momentum). A gap can be opened simply by tuning on a fermion bilinear term $\sum_{i} (c_{iA}^\dagger c_{iB}+\text{h.c.})$ in the Hamiltonian, which preserves the full $\U(1)\times(\dsZ\rtimes\dsZ_2)$ symmetry. This is the familiar band hybridization mechanism to open a band gap in a charge-compensated Fermi liquid, which drives a metal to a band insulator without breaking symmetry.

However, we are more interested in the non-trivial case when the fermions carry different charges $q_A\neq q_B$. For example, let us consider the case of $q_A=1$ and $q_B=3$, then the anomaly cancellation condition \eqnref{eq: anomaly cancellation} requires $\nu_A=-3\nu_B$, i.e.~the electron-like Fermi volume in the $A$ band must be \emph{three times} as large as the hole-like Fermi volume in the $B$ band to cancel the Fermi surface anomaly. Defining the fermion operators $c_{kA}, c_{kB}$ in the momentum space by the Fourier transformation
\eq{\label{eq: def ck}
c_{kA}=\sum_{i}c_{iA}\e^{-\ii k i}, \quad c_{kB}=\sum_{i}c_{iB}\e^{-\ii k i},}
the desired band structure can be effectively described by the following band Hamiltonian (suppressing the interaction for now)
\eq{H=\sum_{k}(c_{kA}^\dagger \epsilon_{kA}c_{kA}+c_{kB}^\dagger \epsilon_{kB}c_{kB}),}
with the band dispersions (see \figref{fig: dispersion}(b))
\eq{\epsilon_{kA}=\frac{k^2-(3k_F)^2}{2m_A},\quad \epsilon_{kB}=-\frac{k^2-k_F^2}{2m_B}.}
Here we assume $m_A,m_B>0$. The Fermi momentum $k_F=|\nu_B|\pi$ is set by the filling $|\nu_B|$ which is typically an \emph{irrational} number (without fine-tuning). The key feature is that the Fermi momenta of the $A$ and $B$ energy bands must have a $3:1$ ratio that matches the inverse charge ratio $(q_A/q_B)^{-1}$ precisely. In this case, the energy band hybridization is forbidden by the charge $\U(1)$ symmetry as the two bands now carry different charges. Even if the band hybridization is spontaneously generated at the price of breaking the $\U(1)$ symmetry, it does not gap the Fermi liquid because the Fermi surfaces of the two bands are no longer nested at the Fermi level, such that the band hybridization will only create some avoided energy band crossing below the Fermi level. Then the system remains metallic because the (upper) hybridized band still crosses the Fermi level.

One can show that it is impossible to symmetrically gap the Fermi liquid by any fermion \emph{bilinear} terms in this charge-compensated two-band system with $q_A=1$ and $q_B=3$, even if the Fermi surface anomaly has already been canceled by the charge-compensated filling $\nu_A=-3\nu_B$. Although the anomaly vanishes (i.e.~there is no obstruction towards gapping in principle), the symmetry is still restrictive enough to forbid any fermion bilinear gapping term, such that the only possible gapping mechanism rests on 
non-perturbative fermion interaction effects. 

To see this, we can single out the low-energy chiral fermions near the four Fermi points: 
\eqs{\label{eq: def ca}
c_{AR}=c_{(3k_F)A},&\quad 
c_{BR}=c_{(-k_F) B},\\
c_{BL}=c_{(k_F)B},&\quad
c_{AL}=c_{(-3k_F)A},} where $A,B$ label the bands that they originated from and $L,R$ label their chiralities (i.e.~left- or right-moving), according to \figref{fig: dispersion}(b). Similar to \eqnref{eq: L FL},
the low-energy effective Lagrangian density reads
\eq{\scL=\sum_a c_a^\dagger(\ii\partial_t + v_a \ii\partial_x)c_a,}
where the index $a$ sums over the four Fermi point labels $AR$, $BR$, $BL$, $AL$. Here $v_a$ denotes the Fermi velocity near the Fermi point $a$.

The original $\U(1)\times \dsZ$ symmetry at the lattice fermion level reduces to the emergent $\U(1)_V\times_{\dsZ_2^F}\U(1)_A$ symmetry
for the low-energy chiral fermions $c_a$ (see \appref{app: symm} for more explanations)
\eqs{\label{eq: ca symm}
\U(1)\Rightarrow\U(1)_V&: c_a\to \e^{\ii q_a^V\phi_V}c_a,\\
\dsZ\Rightarrow\U(1)_A&: c_a\to \e^{\ii q_a^A\phi_A}c_a.\\
}
\tabref{tab: 3450} summarizes their charge assignment under $\U(1)_V$ and $\U(1)_A$, where the vector $\U(1)_V$ symmetry is just the charge $\U(1)$ symmetry and the axial $\U(1)_A$ symmetry is an emergent symmetry corresponding to the lattice translation symmetry $\dsZ$. Alternatively, they can be recombined into the $\U(1)_{\frac{3V+A}{2}}\times\U(1)_{\frac{3V-A}{2}}$ symmetry, such that it becomes obvious that all fermion bilinear back-scattering terms (either the Dirac mass $c_a^\dagger c_b$ or the Majorana mass $c_a c_b$ for $a\neq b$ and $a,b\in\{AR, BR, BL,   AL\}$)  are forbidden by the symmetry because they are all charged non-trivially under the $\U(1)_{\frac{3V+A}{2}}\times\U(1)_{\frac{3V-A}{2}}$ symmetry due to the distinct charge assignment to every chiral fermion. Given this situation, the only hope to gap the Fermi liquid is to evoke the SMG mechanism that generates the mass for all chiral fermions by non-perturbative multi-fermion interactions.

\begin{table}[htp]
\caption{Charge assignments of low-energy fermions. See also the model in \cite{Wang2207.14813} on the same 
charge assignments.}
\begin{center}
\begin{tabular}{clcccc}
fermion & chirality & $\U(1)_V$ & $\U(1)_A$ & $\U(1)_{\frac{3V+A}{2}}$ & $\U(1)_{\frac{3V-A}{2}}$ \vspace{2pt}\\
$c_a$ & $\sgn v_a$ & $q_a^V$ & $q_a^A$ & $\frac{1}{2}(3q_a^V+q_a^A)$ & $\frac{1}{2}(3q_a^V-q_a^A)$ \\
\hline
$c_{AR}$ & {$-1$ (left)} & $1$ & $3$ & $3$ & $0$\\
$c_{BR}$ & {$-1$ (left)} & $3$ & $-1$ & $4$ & $5$\\
$c_{BL}$ & {$+1$ (right)} & $3$ & $1$ & $5$ & $4$\\
$c_{AL}$ & {$+1$ (right)} & $1$ & $-3$ & $0$ & $3$\\
\end{tabular}
\end{center}
\label{tab: 3450}
\end{table}

\subsection{SMG Interaction and RG Analysis}

It is worth mentioning that the charge-compensated two-band model with $q_A=1$ and $q_B=3$ essentially regularizes the 3-4-5-0 chiral fermion model \cite{Bhattacharyahep-lat/0605003, Giedthep-lat/0701004} on a pristine 1D lattice (without introducing any compact extra dimensions). The emergent $\U(1)_{\frac{3V\pm A}{2}}$ symmetries act as the lattice translations decorated by appropriate internal $\U(1)$ rotations, described by the following $\dsZ$ symmetry groups (parameterized by integer $n\in\dsZ$) at the lattice level: (see \appref{app: symm} for derivation)
\eq{\label{eq: 3450 symm}
\dsZ\text{ (for $\tfrac{3V\pm A}{2}$)}:\left\{\begin{array}{l}
c_{iA}\to \e^{\pm \ii 3 k_F n}c_{(i+n)A}, \\
c_{iB}\to \e^{\pm\ii 9 k_F n}c_{(i+n)B}.
\end{array}\right.}
The 3-4-5-0 model is a toy model for studying the long-standing problem: the lattice regularization of the chiral fermion theory in high-energy physics \cite{Nielsen1981a, Nielsen1981b, Nielsen1981NoGo, Swift1984, Eichten1986Chiral, KaplanRLB1992, Bankshep-lat/9204017, Montvayhep-lat/9205023, Wang1809.11171}. Many variants of the model 
are studied in the lattice community \cite{Poppitz1003.5896, Chen1211.6947}. This model is anomaly-free 
--- {perturbative local gauge anomaly free within any linear combination of the $\U(1)_V \times_{\dsZ_2^F} \U(1)_A$ checked by the Adler-Bell-Jackiw method \cite{Adler1969, BellJackiw1969}, perturbative local gravitational anomaly free because of the zero chiral central charge $c_L-c_R=0$, 
also nonperturbative global anomaly free from any gauge or gravitational fields checked by the cobordism \cite{WanWang2018bns1812.11967}.} 
However, it is known much later that symmetric gapping can only be achieved by minimally six-fermion interactions 
among the four flavors of 3-4-5-0 fermions. The SMG interaction was first proposed by Wang and Wen \cite{Wang1307.7480, Wang1807.05998}, which was later discussed by Tong \cite{Tong2104.03997} and only recently verified by the density matrix renormalization group (DMRG) \cite{WhitePRL1992, Schollwockcond-mat/0409292} numerical simulation in \refcite{Zeng2202.12355}.

Given the existing knowledge about the SMG interaction in the 3-4-5-0 chiral fermion model, we can map the Wang-Wen interaction \cite{Wang1307.7480, Wang1807.05998} back to our lattice model following the correspondence listed in \tabref{tab: 3450}, which gives us the following SMG interaction (see \appref{app: wangwen} for more details)
\eq{\label{eq: H two-band interaction}
H_\text{int}=g\sum_{i}c_{(i-1)B}^\dagger c_{(i-1)A} c_{iB}c_{iA}c_{(i+1)B}^\dagger c_{(i+1)A}+\text{h.c.}.}
This is a six-fermion interaction across three adjacent sites on the 1D lattice. It describes the process that first annihilates both $A$- and $B$-type fermions on the center site (which annihilates four units of charges on the site $i$) and then separately converts $A$-type fermions to $B$-type fermions on the two adjacent sites 
(which creates two units of charges 
on each of the site $i-1$ and $i+1$), such that the {$\U(1)_V$} charge is conserved. The interaction is also manifestly translation and reflection symmetric, so the full {$\U(1)_V\times(\dsZ\rtimes\dsZ_2)$} symmetry is preserved by the interaction as expected. With this interaction, we claim that the lattice model \eqnref{eq: H two-band} will exhibit an (ersatz) Fermi liquid to SMG insulator transition when the interaction strength $g$ exceeds a  finite critical value $g_c$. 

\begin{figure}[htbp]
\begin{center}
\includegraphics[scale=0.65]{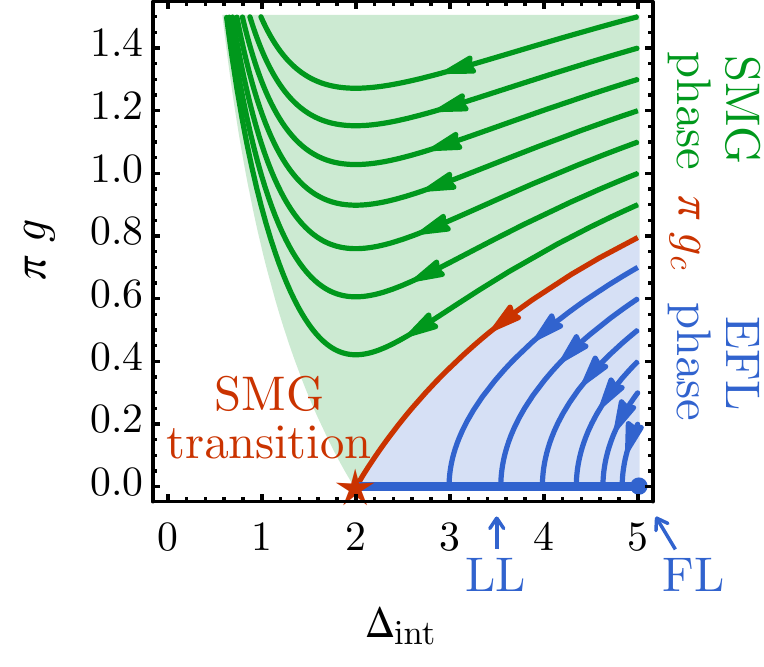}
\caption{The RG flow of the coupling $g$ and the scaling dimension $\Delta_\text{int}$ of the SMG interaction. The abbreviations stand for the following terminology: 
SMG for symmetric mass generation, FL for Fermi liquid, EFL for ersatz Fermi liquid, LL for Luttinger liquid. }
\label{fig: RG flow}
\end{center}
\end{figure}

To show that the proposed interaction \eqnref{eq: H two-band interaction} indeed drives the Fermi liquid to a gapped interacting insulator, we bosonize \cite{Luttinger1963, Fishercond-mat/9610037} the fermion operator $c_{a}\sim{:\,}\e^{\ii\varphi_a}{\,:}$ (with $a\in\{AR, BR, BL, AL\}$) and cast the lattice model to an effective Luttinger liquid theory, described by the following Lagrangian density
\eqs{\label{eq: LL theory}
\scL =& \frac{1}{4\pi}(\partial_t\varphi^\intercal K\partial_x\varphi-\partial_x\varphi^\intercal V\partial_x\varphi)\\
&+\sum_{\alpha=1,2} g_{\alpha}\cos(l_{\alpha}^\intercal\varphi),}
where $\varphi=(\varphi_{AR},\varphi_{BR},\varphi_{BL},\varphi_{AL})^\intercal$ are compact scalar bosons. The $K$ matrix and the $l_\alpha$ vectors are given by
\eq{
K=\smat{1&0&0&0\\0&1&0&0\\0&0&-1&0\\0&0&0&-1},\quad l_1=\smat{1\\-2\\1\\2},\quad
l_2=\smat{2\\1\\-2\\1}.}
The six-fermion interaction $H_\text{int}$ in \eqnref{eq: H two-band interaction} translates to the cosine terms $g_1$ and $g_2$ in the Luttinger liquid theory in \eqnref{eq: LL theory}, with $g_1=g_2=g$ enforced by the $\dsZ_2$ reflection symmetry. The RG flow in the log energy scale $\ell=-\ln\Lambda$ is given by \cite{Kosterlitz1974, Jose1977RG}
\eq{
\frac{\dd g}{\dd \ell}=(2-\Delta_\text{int})g,\quad
\frac{\dd \Delta_\text{int}^{-1}}{\dd \ell}=\pi^2g^2,}
where $\Delta_\text{int}$ is the scaling dimension of the SMG interaction. The RG flow diagram is shown in \figref{fig: RG flow}.

At the Fermi liquid (FL) fixed-point, we have $\Delta_\text{int}=\frac{1}{2}l_\alpha^\intercal l_\alpha=5>2$, meaning that the SMG interaction is perturbatively irrelevant. If the bare coupling $g$ (the interaction strength at the lattice scale) is weak ($g<g_c$), it will just flow to zero and disappear in the IR theory. However, the scaling dimensions of all operators will be renormalized as the coupling $g$ flows toward zero. Therefore, the FL fixed-point will be deformed into the Luttinger liquid (LL) fixed-line, along which the fermion quasi-particle is no longer well-defined, but the system remains gapless. Despite the different dynamical properties, the LL still preserves all the kinematic properties (e.g.~emergent symmetries and anomalies) as the FL, which can be unified under the concept of \emph{ersatz Fermi liquid} (EFL) \cite{Else2007.07896}. 

If the bare coupling $g$ is strong enough ($g>g_c$), the scaling dimension $\Delta_\text{int}$ can be reduced to $\Delta_\text{int}<2$ such that the SMG interaction becomes relevant and flows strong. As the cosine term in \eqnref{eq: LL theory} gets strong, the corresponding vertex operators $\e^{\ii l_\alpha^\intercal\varphi}$ ($\alpha=1,2$) condense. Any other operators that braid non-trivially with the condensed operators will be gapped, which includes all the fermion operators. Therefore the system enters the SMG insulating phase with all fermion excitations gapped without breaking the $\U(1)\times(\dsZ\rtimes\dsZ_2)$ symmetry. This has been confirmed by the DMRG simulation in \refcite{Zeng2202.12355} for a related model using the domain wall fermion construction, where it has been verified that the fermion two-point function indeed decays exponentially in the SMG phase --- a direct piece of evidence for the gap generation. On the lattice level, this corresponds to condensing the six-fermion bound state by developing the ground state expectation value of $\langle c_{(i-1)B}^\dagger c_{(i-1)A} c_{iB}c_{iA}c_{(i+1)B}^\dagger c_{(i+1)A}\rangle\neq 0$. So the gapping is achieved by the multi-fermion condensation (involving more than two fermions), which is distinct from the fermion bilinear condensation in the conventional gapping mechanisms of Fermi liquids (such as the band hybridization or Cooper pairing mechanisms). 

The RG analysis also indicates that the ersatz Fermi liquid to SMG insulator transition (at $g=g_c$) is of the Berezinskii-Kosterlitz-Thouless (BKT) \cite{Berezinskii1971, Berezinskii1972, KosterlitzThouless1973} transition universality in (1+1)D.

The above analysis established the Fermi surface SMG phenomenon in the lattice model \eqnref{eq: H two-band} (equipped with the gapping interaction \eqnref{eq: H two-band interaction}). The significance of this lattice model is that it provides a pristine 1D lattice regularization of the 3-4-5-0 chiral fermion model by using lattice translation to realize the axial $\U(1)_A$ symmetry at low energy. This construction does not require the introduction of a (2+1)D bulk to realize the chiral fermions as boundary modes, as in the domain wall fermion constructions \cite{Wang1807.05998, Wang1809.11171, Zeng2202.12355}. Such pristine 1D lattice regularization is advantageous for the numerical simulation of chiral fermions, as the model contains no redundant bulk (or mirror) fermions, such that the computational resources can be used more efficiently. We will leave the numerical exploration of this model to future research.

\section{Fermi Surface SMG in (2+1)D}\label{sec:2+1}

\subsection{(2+1)D Fermi Liquid and Fermi Surface Anomaly}

Given the example of Fermi surface SMG in (1+1)D, we would like to further explore similar physics in higher dimensions. The most important low-energy features of a (2+1)D Fermi liquid are the gapless fermions on its 1D Fermi surface. Suppose we parameterize the 1D Fermi surface $\vect{k}_F(\theta)\in\partial\scV_F$ by a continuous and periodic parameter $\theta$, such that $\vect{k}_F(\theta+2\pi)=\vect{k}_F(\theta)$ (where we do not require $\theta$ to literally represent the geometrical angle, as the Fermi surface may not be a perfect circle in general). The fermions $c_\theta$ on the Fermi surface have an emergent symmetry described by the loop group of $\U(1)$ \cite{Else2007.07896,Else2010.10523}, denoted as $\LU(1)$, under which
\eq{
\LU(1):c_\theta\to \e^{\ii\phi(\theta)} c_\theta,}
where the $\U(1)$ phase factor $\e^{\ii\phi(\theta)}$ is a smooth function of $\theta$ with the periodicity $\e^{\ii\phi(\theta+2\pi)}=\e^{\ii\phi(\theta)}$. Both the (global) charge $\U(1)$ and the translation symmetries $\dsR^2$ are subgroups of $\LU(1)$:
\eq{
\U(1):c_\theta\to\e^{\ii q\phi}c_\theta,\quad\dsR^2:c_\theta\to\e^{\ii\vect{a}\cdot\vect{k}_F(\theta)}c_\theta,}
assuming the fermions $c_\theta$ carry charge $q$ under the global $\U(1)$ symmetry and are translated by the vector $\vect{a}\in\dsR^2$.

The presence of the Fermi surface causes a mixed anomaly between the $\U(1)$ and translation symmetries \cite{Wen2101.08772}, which is characterized by the anomaly index
\eq{\label{eq: 2D anomaly}
\frac{q}{2(2\pi)^2}\oint\dd\theta (\vect{k}_F\times\partial_\theta \vect{k}_F)_3 =\frac{q \scV_F}{(2\pi)^2}=q\nu,}
where $\scV_F$ stands for the Fermi volume in the momentum space, and $\nu$ is the filling factor. If the Fermi surface anomaly is non-vanishing, it is impossible to trivially gap out the Fermi liquid without breaking any symmetry or developing any topological order. The Fermi surface SMG is only possible if the Fermi liquid system contains multiple Fermi surfaces of opposite anomaly indices, such that their anomalies cancel as a whole.

\subsection{Kagome-Triangular Lattice Model}

\begin{figure}[htbp]
\begin{center}
\includegraphics[scale=0.65]{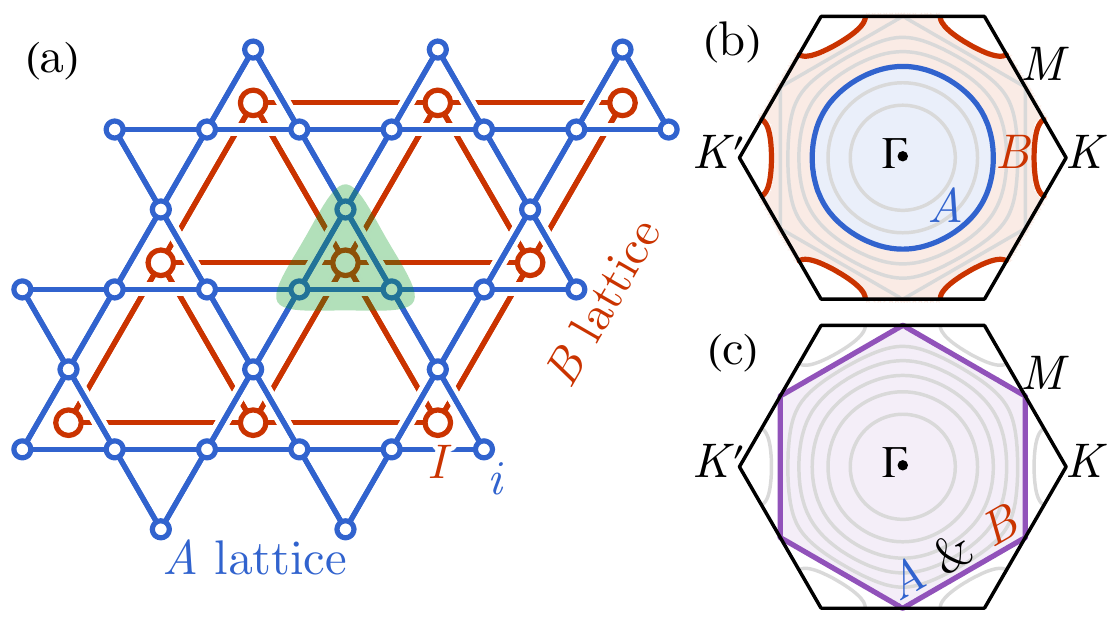}
\caption{(a) Overlapping Kagome ($A$) and triangular ($B$) lattices. The green triangle marks out the unit cell. $A$-type (in blue) and $B$-type (in red) Fermi surfaces (b) at a general filling such as $\nu_A=3/8$ and $\nu_B=7/8(=-1/8)$, or (c) at a special filling $\nu_A=\nu_B=3/4(=-1/4)$ where the Fermi surfaces coincide.}
\label{fig: 2D lattice}
\end{center}
\end{figure}

We present a concrete lattice model to demonstrate the Fermi surface SMG in (2+1)D. Consider two types of \emph{spinless} fermions labeled by $A$ and $B$ that are charged under a global $\U(1)$ symmetry with charges $q_A=1$ and $q_B=3$, respectively. The $A$-type (or $B$-type) fermion is defined on a Kagome (or triangular) lattice. As depicted in \figref{fig: 2D lattice}(a), the Kagome and the triangular lattices lie on top of each other, with the site $I$ of the triangular lattice aligned with the upper triangle $\triangle_I$ on the Kagome lattice. We will use the lower-case letters $i, j$ (or the upper-case letters $I, J$) to label the Kagome (or the triangular) lattice sites. 

The lattice model is described by following the Hamiltonian
\eqs{\label{eq: 2D model}
H&=H_A+H_B+\hcf,\\
H_A&=-t_A\sum_{\langle ij\rangle}(c_i^\dagger c_j+\text{h.c.})-\mu_A\sum_{i}c_i^\dagger c_i,\\
H_B&=-t_B\sum_{\langle IJ\rangle}(c_I^\dagger c_J+\text{h.c.})-\mu_B\sum_{I}c_I^\dagger c_I,\\
\hcf&=-g\sum_{I|ijk\in\triangle_I}(c_I^\dagger c_ic_jc_k+\text{h.c.}),}
where $\langle ij\rangle$ (or $\langle IJ\rangle$) denotes the nearest neighboring link on the $A$ (or $B$) lattices and $ijk\in\triangle_I$ stands for the three $A$-sites $i,j,k$ at the vertices of the upper triangle surrounding the $B$-site labeled by $I$. The model has a $\U(1)$ symmetry that acts as
\eq{\label{eq: U(1) 2D}\U(1): c_i\to\e^{\ii\phi}c_i,\quad c_I\to \e^{\ii 3\phi}c_I.}
The Hamiltonian in \eqnref{eq: 2D model} preserves the internal $\U(1)$ symmetry and all symmetries of the Kagome-triangular lattice (most importantly, the lattice translation symmetry). 

The model \eqnref{eq: 2D model} describes the two types of fermions hopping separately on their corresponding lattices. Because every unit cell contains four sites (three from the Kagome lattice and one from the triangle lattice), the hopping model will give rise to four energy bands (three bands for $A$-type fermions and one band for $B$-type fermions). The chemical potentials $\mu_A$ and $\mu_B$ are adjusted to ensure the desired filling of these fermions. We will focus on a simple case when only the lowest $A$-type (Kagome lattice) bands and the $B$-type (triangular lattice) bands are filled by filling fractions $\nu_A$ and $\nu_B$ respectively, such that the Fermi surface only involves \emph{two} of the four bands.

The $A$-type and $B$-type fermions are coupled together only through a four-fermion interaction $\hcf$ in \eqnref{eq: 2D model} that fuses three $A$-type (charge-1) fermions to one $B$-type (charge-3) fermions (and vice versa) within each unit cell. We will call it a \emph{charge fusion} (CF) interaction. The CF interaction breaks the separate $U(1)$ charge conservation laws for $A$-type and $B$-type fermions in the hopping model to a joint $\U(1)$ charge conservation, associated with the symmetry action in \eqnref{eq: U(1) 2D}. Similar interactions also appear in a recent study \cite{Lian2208.10509} of quantum breakdown. 

Without interaction ($g=0$), the system is in a Fermi liquid phase. According to \eqnref{eq: 2D anomaly}, the Fermi surface anomaly cancellation condition requires
\eq{q_A\nu_A+q_B\nu_B=0\mod 1.}
Given the charge assignment of $q_A=1$ and $q_B=3$, it requires $\nu_A=-3\nu_B$. There is no further requirement on the choice of $\nu_A$ itself. With a generic choice of filling (assuming $\nu_A<3/4$) as in \figref{fig: 2D lattice}(b), the $A$-type fermions (on the Kagome lattice) will form an electron-like Fermi surface, whose Fermi volume is three times as large as that of the hole-like Fermi surfaces formed by the $B$-type fermions (on the triangular lattice). Although the Fermi liquid has a vanishing Fermi surface anomaly, the charge $\U(1)$ and the lattice translation symmetries are still restrictive enough to forbid any gap opening on the free-fermion level. For example, any pairing (charge-2e superconducting) gap will break the $\U(1)$ symmetry. The only possibility to gap the Fermi liquid relies on the fermion interaction. 

We claim that the charge fusion interaction $\hcf$ in \eqnref{eq: 2D model} is a valid SMG interaction that drives the Fermi liquid into a trivially gaped insulator without breaking symmetry (or developing any topological order). To see this, we go to the strong coupling limit by taking $g\to\infty$. Of course, the chemical potentials $\mu_A,\mu_B$ must increase correspondingly to keep the fermion fillings fixed. The model Hamiltonian decouples to each unit cell in the strong coupling limit
\eqs{
H=\sum_{I|ijk\in\triangle_I}&-\mu_A(n_i+n_j+n_k)-\mu_B n_I\\
&-g(c_I^\dagger c_ic_jc_k+\text{h.c.}),}
where $n_i=c_i^\dagger c_i$ (and $n_I=c_I^\dagger c_I$) denotes the fermion number operator. Within each unit cell, there are only two relevant states $\ket{1110}$ and $\ket{0001}$ (in the Fock state basis $\ket{n_in_jn_kn_I}$) acted by the Hamiltonian. Their hybridization will produce the ground state in each unit cell. The full-system ground state will be the following direct product state
\eq{
\ket{\text{SMG}}=\bigotimes_{I}\big(\sqrt{p}\ket{1110}+\sqrt{1-p}\ket{0001}\big)_I,}
where $p=\frac{1}{2}\big(1+\frac{-3\mu_A+\mu_B}{\sqrt{(-3\mu_A+\mu_B)^2+4g^2}}\big)$ is the probability to observe the $\ket{1110}$ state in the unit cell, which is tunable by adjusting $\mu_A,\mu_B$ relative to $g$. The fermion fillings (per unit cell) in the ground state $\ket{\text{SMG}}$ will be 
\eq{\nu_A=3p,\quad \nu_B=1-p=-p \mod 1,}
which automatically satisfies the anomaly cancellation condition $\nu_A=-3\nu_B$ (as it should be). The ground state $\ket{\text{SMG}}$ is non-degenerated and gapped from all excited states (with a gap of the order $g$). It also preserves the charge $\U(1)$ and all the lattice symmetries and does not have topological order. Therefore, we have explicitly shown that the system ends up in the SMG insulator phase as $g\to \infty$. As a gapped phase, we expect it to be stable against perturbations (such as the hopping terms $t_A,t_B$) over a finite region in the parameter space. The SMG phase is a strongly interacting insulating phase, which has no correspondence in the free-fermion picture.

Established the Fermi liquid (metallic) phase at $g=0$ and the SMG insulator phase at $g\to\infty$, there must be an SMG transition (an interaction-driven metal-insulator transition) at some intermediate coupling strength $g_c$. However, the nature of the transition is still an open question, which we will leave for future numerical study. In the following, we will only analyze the SMG transition at a special filling: $\nu_A=\nu_B=3/4$, where the Fermi surfaces coincide precisely and take the perfect hexagon shapes as shown in \figref{fig: 2D lattice}(c). This allows us to gain some analytic control of the problem.

\subsection{RG Analysis of the SMG Transition}
In this subsection, we analyze the interaction effect in \eqnref{eq: 2D model} when the filling is $\nu_A=\nu_B=3/4$. In this case, the Fermi surface contains three Van Hove singularities (VHSs) (or three hot spots) at three distinct $M$ points as shown in \figref{fig: 2D lattice}(c). We can investigate the interaction effects using the hot-spot renormalization group (RG) method at the one-loop level \cite{Furukawacond-mat/9806159, Raghu1009.3600, Nandkishore1107.1903, Isobe1805.06449, Lin1901.00500, Lin2008.05485, Park2104.08425, You2109.04669}. The hot-spot RG assumes that the low-energy physics emerges from the correlated effects of the fermions near the VHSs, because the density of states diverges near the VHSs, leading to the highest instability towards ordering.

Under RG, the charge fusion interaction $\hcf$ will generate two types of density-density interactions at the one-loop level, namely, $\hda = \sum_{i,j} n_i n_j$ and $\hdb =\sum_{I,i} n_I n_i$ as well as other (less important) exchange interactions. These density-density interactions are more important in the sense that they will in turn contribute to the correction of $\hcf$. Therefore, we should include $\hcf,\hda,\hdb$ altogether in the RG analysis and study the RG flow jointly. 

To proceed, we transform the interactions into the momentum space. The fermion operators are labeled by the flavor index $S=A, B$ and the hot-spot index $\alpha=1,2,3$ (referring to the three different VHSs). We note that $\hcf$ would vanish if it is naively restricted to the hot spots because the momentum conservation requires multiple $A$-type fermion operators to appear on the same hot spot, which violates the Pauli exclusion principle of fermions. So we need to introduce point splitting in the momentum space around each hot spot. Our strategy is to further split the $A$-type fermion into three modes $A_s$ labeled by $s=1,2,3$, and define the interaction,
\begin{align}
    &\hcf = g_\text{rs} \sum_\alpha \epsilon^{ijk} c_{B \alpha}^\dagger c_{A_i \alpha}c_{A_j \alpha}c_{A_k \alpha} \nonumber \\
    &+g_\text{rt} \sum_{\alpha \ne \beta} \epsilon^{ijk} c_{B \alpha}^\dagger c_{A_i \alpha}c_{A_j \beta}c_{A_k \beta}  +\text{h.c.}
\end{align}
$g_\text{rs}$ and $g_\text{rt}$ are the CF interaction decomposed into different momentum transfer channels: the intra-hot-spot scattering $g_\text{rs}$ and the inter-hot-spot scattering $g_\text{rt}$.

\begin{figure}[htbp]
    \centering
    \includegraphics[scale=0.65]{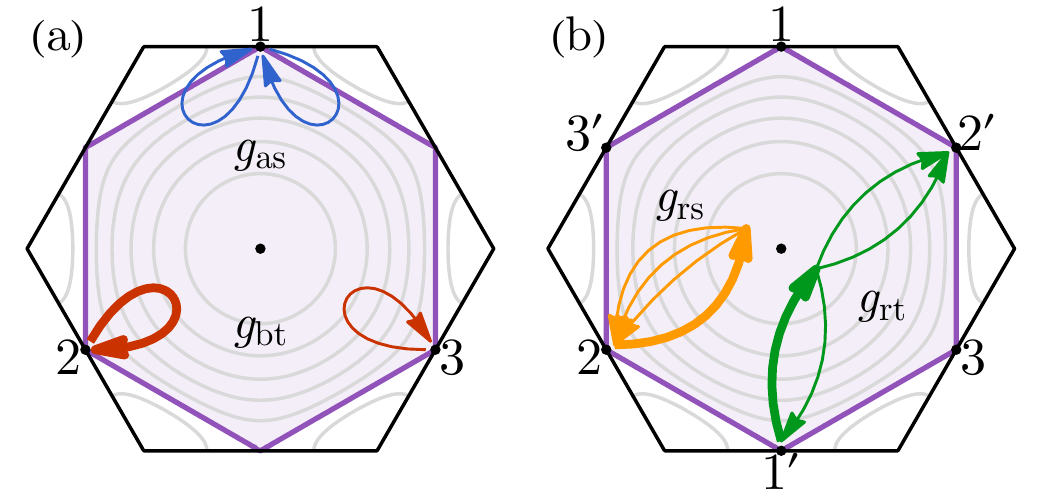}
    \caption{Scattering of fermions between Van Hove singularities 
    (VHSs) by (a) density-density interactions $g_\text{bt}$ (red), $g_\text{as}$ (blue) and (b)  non-vanishing processes $g_\text{rs}$ (yellow), $g_\text{rt}$ (green) of $\hcf$. Thin (or thick) arrows correspond to $A$-type (or $B$-type) fermions.}
    \label{fig: process}
\end{figure}

These CF interactions receive corrections from the following density-density interactions at the one-loop level,
\begin{align}
    &\hda+\hdb =g_\text{as} \sum_{\alpha, s t}  n_{A_{s}\alpha}n_{A_{t}\alpha}+(A_{s}\leftrightarrow A_{t}) \nonumber\\
    &+g_\text{bt} \sum_{\alpha\ne \beta,s}  n_{B\alpha}n_{A_s\beta}+(A_s\leftrightarrow B)+\text{h.c.} +...
\end{align}
where $...$ refers to the other interactions that are decoupled from $g_\text{rs},g_\text{rt},g_\text{as},g_\text{bt}$ in the RG equations. The scattering processes of these four important interactions are illustrated in \figref{fig: process}. The complete set of all possible interactions is presented in \appref{app:fullrge}. 

We derive the RG equations based on the systematic approach developed in \refcite{Lu2206.01213}. Since we are interested in the flow of $\hcf$, the relevant part of the RG equations reads,
\begin{align} \label{eq: RGequ}
    &\frac{d g_\text{bt} }{d \ell} = 2 d_0 d_{\text{AB}} g_\text{bt}^2,\quad \frac{d g_\text{as}}{d\ell} = -2 g_\text{as}^2,  \\
    & \frac{d g_\text{rs} }{d\ell} = -6 g_\text{as} g_\text{rs}, \quad  \frac{d g_\text{rt}}{d \ell} = 4d_0 d_{\text{AB}} g_\text{bt}g_\text{rt} -2g_\text{as}g_\text{rt}. \nonumber
\end{align}
where the RG parameter is defined by the Cooper-pairing susceptibility of $A$-type fermions $\ell=\chi_{pp,\text{AA}}(\vect{k}=0,E)\sim \nu_0 \log^2(\Lambda/E)$, in which $\nu_0\log(\Lambda/E)$ is the diverging density of states at the VHS, $E$ is the running energy scale and $\Lambda$ is the high energy cutoff. $d_0=d\chi_{ph,\text{AA}}(\vect{Q})/d\ell\leq1$ is the nesting parameter of $A$-type fermions, which saturates to one in the perfectly nested limit ($d_0\to 1$). In our case, different VHSs are half-nested (only one of the two crossing Fermi surfaces is perfectly nested between every pair of different VHSs), so $d_0=1/2$ is a suitable estimation. Similarly, we define $d_{\text{AB}}=d\chi_{pp,\text{AB}}(\vect{0})/d\ell$, which depends on the energies of $A$ and $B$-type fermions near the VHS. The full RG equations and details are listed in \appref{app:fullrge}. 


According to the one-loop RG equations, if the density-density interactions $g_\text{bt},g_\text{as}$ are initially zero, then the CF interactions $g_\text{rs},g_\text{rt}$ remain marginal along the RG flow. However, if we turn on small density-density interactions $g_\text{bt}, g_\text{as}$ with correct signs ($g_\text{bt}>0$ or $g_\text{as}<0$), the charge fusion interactions $g_\text{rs},g_\text{rt}$ will be marginally relevant. The solutions of RG equations \eqnref{eq: RGequ} are
\begin{align}\label{eq: RGsol}
    &g_\text{bt}(\ell) = \frac{g_\text{bt}(0)}{1-2d_0 d_{\text{AB}} g_\text{bt}(0) \ell},\quad  g_\text{as}(\ell) = \frac{g_\text{as}(0)}{1+2 g_\text{as}(0) \ell}, \nonumber \\
    & g_\text{rs}(\ell) = \frac{g_\text{rs}(0)}{(1+2 g_\text{as}(0) \ell)^3}, \nonumber\\
    &g_\text{rt}(\ell) = \frac{g_\text{rt}(0)}{(1+2 g_\text{as}(0) \ell)(1-2d_0 d_{\text{AB}} g_\text{bt}(0) \ell)^2} .
\end{align}
As the RG parameter $\ell$ increases under the RG flow, the coupling strengths can diverge at some critical scale $\ell_c$, when any of the denominators in \eqnref{eq: RGsol} vanish. The critical scale is set by the bare density-density interaction strengths $g_\text{bt}(0)$ and $g_\text{as}(0)$, but the CF interaction strengths $g_\text{rs},g_\text{rt}$ diverge faster than the density-density interactions as the critical scale is approached. Therefore, the RG fixed points are characterized by the behavior of $g_\text{rs},g_\text{rt}$.

\begin{figure}[htbp]
    \centering
    \includegraphics[scale=0.65]{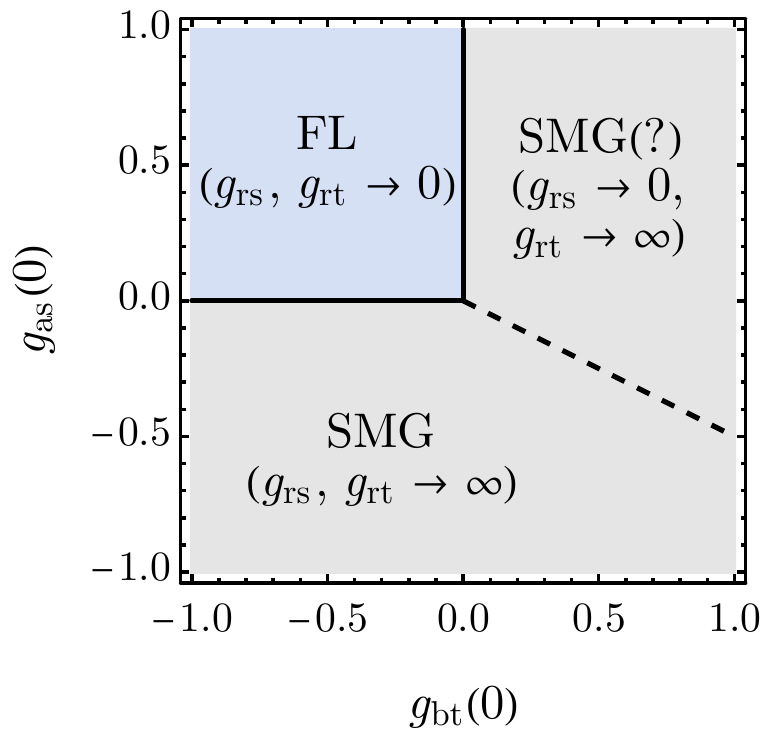}
    \caption{The RG phase diagram with respect to the density-density interactions $g_\text{as}, g_\text{bt}$. In the Fermi liquid (FL) phase, the gapping interaction flows to zero. In the symmetric mass generation (SMG) phase, the gapping interaction flows to infinity.
    }
    \label{fig:phasediagram}
\end{figure}

Depending on the bare density-density interaction strengths $g_\text{as}(0)$ and $g_\text{bt}(0)$, the system can flow towards different RG fixed points as shown \figref{fig:phasediagram}. When $g_\text{as}(0)>0$ and $g_\text{bt}(0)<0$, all interactions flow to zero, which corresponds to the Fermi liquid fixed point. When $g_\text{as}(0)<\min(0, -d_0 d_{\text{AB}} g_\text{bt}(0))$, both charge fusion interactions $g_\text{rs},g_\text{rt}$ flow to infinity, which should correspond to the SMG phase according to the previous lattice model analysis. However, we also find a region in the phase diagram, described by $g_\text{bt}>\max(0,-g_\text{as}/d_0 d_{\text{AB}})$, where $g_\text{rs}\to 0$ and $g_\text{rt}\to \infty$, i.e. flowing towards different limits. We are not sure how to interpret the physical meaning of this RG fixed point. It might still be in the SMG phase as one interaction still flows strong. But it could as well end up in a spontaneous symmetry breaking (SSB) phase that breaks the $\LU(1)$ symmetry since the $A$-type and $B$-type Fermi surfaces have pretty strong nesting instability. This might also be an artifact of the hot-spot RG method, as it does not fully capture all low-energy fermionic degrees of freedom of the Fermi surface. 

Functional RG \cite{Wetterich1993, SalmhoferHonerkamp2001, Dupuis2006.04853,QHWFRG} might provide a better resolution of the Fermi surface and remove the uncertainty in the phase diagram \figref{fig:phasediagram}. A recent study \cite{Gneist2205.12547} has demonstrated the functional RG method in a triangle lattice model with spinless fermions. The same technique might apply to our model as well. However, we will leave such study for future research.

By tuning $g_\text{as}(0)$ across zero on the $g_\text{bt}(0)<0$ side, one can drive a FL to SMG transition. The gapping interaction is marginally relevant at the transition point. According to the solution of the RG equations in \eqnref{eq: RGsol}, the coupling diverges at the critical scale $\ell_c\sim\nu_0\log^2(\Lambda/\Delta_\text{SMG})$ when the denominator $(1+2g_\text{as}(0)l_c)$ vanishes. This implies that the SMG gap $\Delta_\text{SMG}$ (the energy gap between the ground state and the first excited state) opens up as \cite{Sonhep-ph/9812287, Moon1005.0356}
\begin{equation}
\Delta_\text{SMG} \sim \Lambda \exp(-c/\sqrt{g_\text{as}(0) \nu_0}),
\end{equation}
where $\Lambda$ is the UV cutoff energy scale, $\nu_0$ is the coefficient in front of the diverging density of state at the VHS, and $c$ is some non-universal constant.


\section{Summary and Discossion}\label{sec:sum}

In this work, we propose the concept of Fermi surface SMG: a mechanism to gap out Fermi surfaces by non-perturbative interaction effects without breaking the $\LU(1)$ symmetry. This phenomenon can only happen when the Fermi surface anomaly is canceled out in the fermion system. We present (1+1)D and (2+1)D examples of Fermi surface SMG. We expect that the mechanism can generally occur in all dimensions. 

\begin{figure}[htbp]
    \centering
    \includegraphics[scale=0.65]{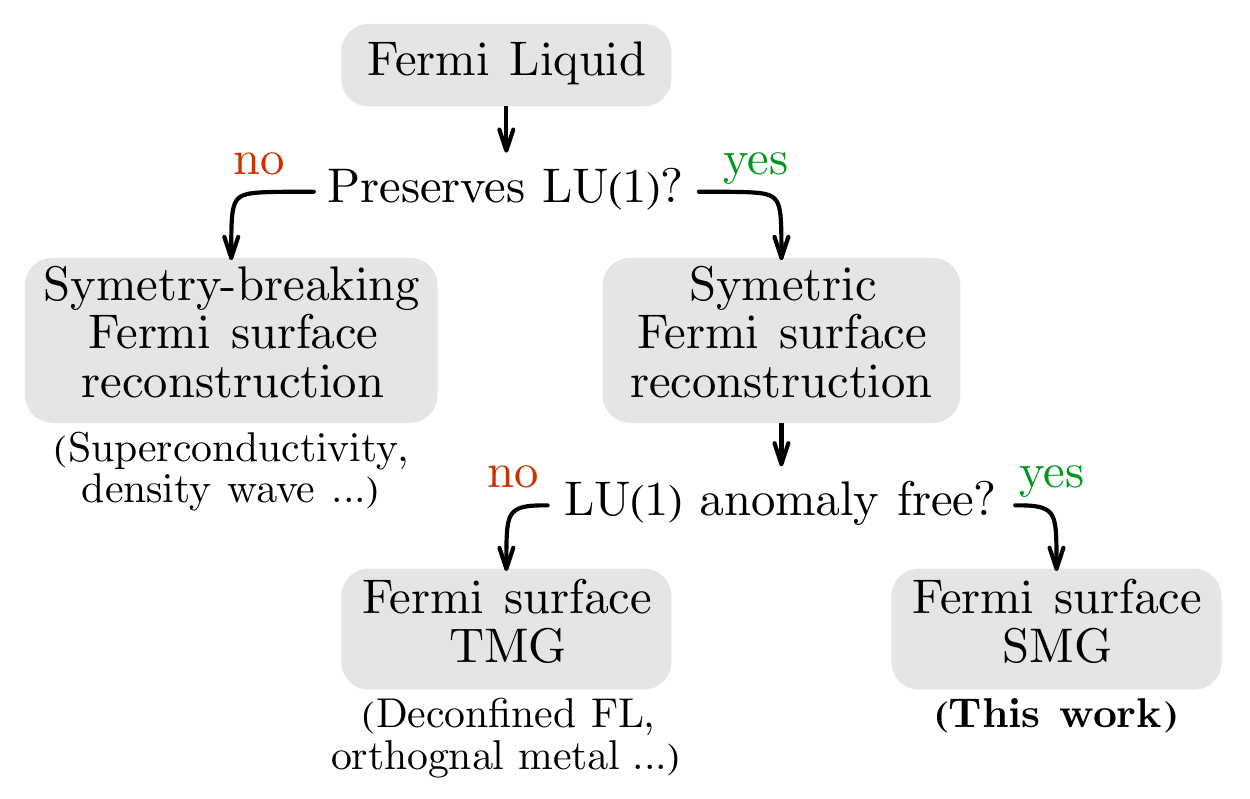}
    \caption{Classification of Fermi surface reconstruction mechanisms, based on the $\LU(1)$ loop group symmetry.}
    \label{fig: tree}
\end{figure}

Fermi surface SMG belongs to a broader class of phenomena, called the symmetric Fermi surface reconstruction (SFSR), as summarised in \figref{fig: tree}. The SFSR is in contrast to the more conventional symmetry-breaking Fermi surface reconstruction, where the Fermi surface is reconstructed (or gapped) by developing spontaneous symmetry-breaking orders. Depending on the cancellation of the Fermi surface anomaly, the SFSR further splits into two classes: the Fermi surface symmetric mass generation (SMG) if the anomaly vanishes, or the Fermi surface topological mass generation (TMG) if the anomaly does not vanish. The former class, the Fermi surface SMG, is the focus of this work. The latter class, the Fermi surface TMG, is also discussed in the literature, where the non-vanishing Fermi surface anomaly is absorbed by an anomalous topological quantum field theory (TQFT), such that the SFSR is achieved by developing the corresponding topological order. This gives rise to deconfined/fractionalized Fermi liquid (FL$^*$) \cite{Senthilcond-mat/0209144, Senthilcond-mat/0305193, Gazit1906.11250} or orthogonal metal \cite{Nandkishore1201.5998, Hohenadler1804.05858, Chen1904.12872}. 
\emph{Symmetry extension} \cite{Wang2018Symmetric} has provided a unified framework, to understand TMG and SMG for bosons or fermions of zero Fermi volume \cite{Tachikawa2017gyf1712.09542, 
Wang2018Tunneling1801.05416,
Guo2020Fermionic1812.11959,
KobayashiOhmoriTachikawa1905.05391,
Prakash2018Unwinding1804.11236, 
Prakash2021Unwinding2011.13921}, where the symmetric gapping can be achieved by extending the symmetry group to lift any gapping obstruction that was otherwise imposed by the symmetry. Similar constructions can be applied to understand SFSR more generally.

Fermi surface SMG deforms an anomaly-free (charge-compensated) Fermi liquid state to a fully gapped product state. Although the resulting SMG gapped state does not have non-trivial features like topological order, the SMG transition from the Fermi liquid phase to the SMG phase can still be quite exotic. The SMG transition of relativistic fermions has been proposed to be a deconfined quantum critical point \cite{You1705.09313, You1711.00863}, where the physical fermion fractionalizes to bosonic and fermionic partons with emergent gauge fluctuations at and only at the critical point. It is conceivable that similar scenarios might apply to the Fermi surface SMG transition as well, where deconfined Fermi liquid (orthogonal metal) could emerge at the critical point. The lattice models presented in this study lay the ground for future theoretical and numerical studies of the exotic SMG transition in these models.

It is also known that the fermion single-particle Green's function has symmetry-protected zeros at zero frequency in the SMG phase \cite{You1403.4938, Catterall1609.08541, Catterall1708.06715, Xu2103.15865}. It will be interesting to investigate further the Green's function structure in the Fermi surface SMG phase. Whether or not the SMG interaction will replace the original Fermi surface (a loop of poles) with a loop of zeros in the Green's function is still an open question to explore.

\begin{acknowledgments}
We thank David Tong, Tin Sulejmanpasic, Max Metlitski, Xiao-Liang Qi, Cenke Xu, Subir Sachdev, John Preskill for inspiring discussions. We acknowledge the workshop ``Paths to Quantum Field Theory 2022'' at  Durham University, where the discussion with David Tong and Tin Sulejmanpasic motivates the authors to think about pristine lattice regularizations of the 3-4-5-0 chiral fermion model. DCL, MZ, and YZY are supported by a startup fund at UCSD. JW is supported by the Center for Mathematical Sciences and Applications at Harvard University and
NSF Grant DMS-1607871 ``Analysis, Geometry and Mathematical Physics.'' 
\end{acknowledgments}

\bibliography{ref}
\clearpage
\appendix
\section{Emergent $\U(1)$ Symmetries in the (1+1)D Two-Band Model}\label{app: symm}
Start from the definition of charge $\U(1)$ (parameterized by a periodic angle $\phi\in[0,2\pi)$) and lattice translation $\dsZ$ (parameterized by an integer $n\in\dsZ$) symmetries as defined in \eqnref{eq: latt symm}
\eqs{
\U(1)&:c_{iA}\to\e^{\ii q_A\phi}c_{iA},\quad c_{iB}\to\e^{\ii q_B\phi}c_{iB};\\
\dsZ&:c_{iA}\to c_{(i+n)A},\quad c_{iB}\to c_{(i+n)B}.}
Follow the definition \eqnref{eq: def ck} of the fermion operators in the momentum space
\eq{
c_{kA}=\sum_{i}c_{iA}\e^{-\ii k i}, \quad c_{kB}=\sum_{i}c_{iB}\e^{-\ii k i},}
where the wave number $k\in[-\pi,\pi)$ is a dimensionless periodic 
variable defined in the first Brillouin zone. (Note: the \emph{dimensionful} momentum $p$ should be related to the \emph{dimensionless} wave number $k$ by $p=\hbar k/a$ with $a$ being the lattice constant and the site coordinate $x\in\dsR$ is related to the site index $i\in \dsZ$ by $x=a i$, such that the Fourier factor $\e^{-\ii p x/\hbar}=\e^{-\ii k i}$ is consistent with the quantum mechanics convention.) It is straightforward to show that the $\U(1)\times\dsZ$ symmetry acts in the momentum space as
\eqs{\label{eq: ck symm} \U(1)&:c_{kA}\to e^{\ii q_A\phi}c_{kA}, c_{kB}\to e^{\ii q_B\phi}c_{kB};\\
\dsZ&:c_{kA}\to e^{\ii kn}c_{kA}, c_{kB}\to e^{\ii k n}c_{kB}.\\
}
Apply these transformations to the low-energy fermion near the four Fermi points. According to \eqnref{eq: def ca}
\eqs{c_{AR}&=c_{(3k_F)A},\\
c_{BR}&=c_{(-k_F) B},\\
c_{BL}&=c_{(k_F)B},\\
c_{AL}&=c_{(-3k_F)A},} 
\eqnref{eq: ck symm} becomes
\eqs{\label{eq: ca symm1}
\U(1)&:\left\{\begin{array}{rcl}
c_{AR} & \to & \e^{\ii q_A\phi}c_{AR},\\
c_{BR} & \to & \e^{\ii q_B\phi}c_{BR},\\
c_{BL} & \to & \e^{\ii q_B\phi}c_{BL},\\
c_{AL} & \to & \e^{\ii q_A\phi}c_{AL};\\
\end{array}\right.\\
\dsZ&:\left\{\begin{array}{rcl}
c_{AR} & \to & \e^{3 \ii  k_F n}c_{AR},\\
c_{BR} & \to & \e^{-\ii k_F n}c_{BR},\\
c_{BL} & \to & \e^{\ii k_F n}c_{BL},\\
c_{AL} & \to & \e^{-3 \ii k_F n}c_{AL}.\\
\end{array}\right.
}
Because $k_F=|\nu_B|\pi$ is \emph{almost always} (i.e., with probability 1) a irrational multiple of $\pi$ (because $|\nu_B|$ is almost always an irrational number without fine tuning), $k_F n\mod 2\pi$ can approach any angle in $[0,2\pi)$ (with $2\pi$ periodicity) as close as we want (given $n\in \dsZ$). This allows us to define two angular variables $\phi_V$ and $\phi_A$, both are periodic in $[0,2\pi)$,
\eq{\label{eq: def phiVA}
\phi_V=\phi,\quad\phi_A=k_F n\mod 2\pi,}
then \eqnref{eq: ca symm1} can be compactly written as
\eqs{\label{eq: ca symm2}
\text{UV symmetry}&\Rightarrow\text{IR symmetry}\\
\U(1)&\Rightarrow\U(1)_V:c_a\to \e^{\ii q_a^V\phi_V} c_a,\\
\dsZ&\Rightarrow\U(1)_A:c_a\to e^{\ii q_a^A \phi_A}c_a,
}
for $a=AR,BR, BL,AL$ enumerating over the four Fermi point labels, together with the charge vectors (given that $q_A=1$ and $q_B=3$):
\eq{
\vect{q}^V=\mat{q_A\\q_B\\q_B\\q_A}=\mat{1\\3\\3\\1},\quad \vect{q}^A=\mat{3\\-1\\1\\-3}.
}
Therefore, the global charge $\U(1)$ symmetry is simply reinterpreted as the $\U(1)_V$ vector symmetry, and the translation symmetry (described by a non-compact $\dsZ$ group) in the UV becomes an emergent $\U(1)_A$ axial symmetry (described by a compact $\U(1)$ group) in the IR. The symmetry transformation in \eqnref{eq: ca symm2} precisely matches \eqnref{eq: ca symm} in the main text with the correct charge assignment as listed in \tabref{tab: 3450}.

Recombining the charge vectors of $\U(1)_V$ and $\U(1)_A$, we can define two alternative emergent $\U(1)$ symmetries, denoted as $\U(1)_{\frac{3V\pm A}{2}}$ with the charge vectors
\eq{\vect{q}^{\frac{3V\pm A}{2}}=\frac{1}{2}(3\vect{q}^V\pm\vect{q}^A),}
as their names implied. More explicitly, the charge vectors match the chiral charge assignements for the 3-4-5-0 fermions:
\eq{\vect{q}^{\frac{3V+ A}{2}}=\mat{3\\4\\5\\0},\quad\vect{q}^{\frac{3V- A}{2}}=\mat{0\\5\\4\\3}.}
The fermions are expected to transform under $\U(1)_{\frac{3V\pm A}{2}}$ as (parameterized by the periodic angles $\phi_\pm\in [0,2\pi)$)
\eq{\label{eq: def 3450 symm}
\U(1)_{\frac{3V\pm A}{2}}:c_a\to \e^{\ii \frac{1}{2}(3q_a^V\pm q_a^A)\phi_\pm}c_a.}
This can be viewed as the combined transformation of $\U(1)_V$ and $\U(1)_A$ with the vector rotation angle $\phi_V$ and the axial rotation angle $\phi_A$ given by
\eq{\phi_V=\frac{3}{2}\phi_\pm, \quad \phi_A=\pm\frac{1}{2}\phi_\pm,}
as can be verified by comparing \eqnref{eq: def 3450 symm} with \eqnref{eq: ca symm2}. Now we can connect these rotation angle back to the original $\U(1)\times\dsZ$ symmetry of the lattice fermions using the relation \eqnref{eq: def phiVA},
\eq{\phi=\frac{3}{2}\phi_\pm,\quad \pm\frac{1}{2}\phi_\pm=k_F n \mod 2\pi.}
Eliminate $\phi_\pm$ from the equations, we obtain the relation
\eq{\phi=\pm 3k_F n\mod 2\pi,}
for the $\U(1)_{\frac{3V\pm A}{2}}$ symmetries. Therefore, in order to reproduce the IR emergent $\U(1)_{\frac{3V\pm A}{2}}$ symmetries, the corresponding  UV symmetries (at the lattice level) must be such implemented that every $n$-step translation should be followed by a charge $\U(1)$ rotation with the rotation angle $\phi=\pm 3k_Fn$. So we establish the following  correspondence between the IR and UV symmetries
\eqs{\label{eq: 3450 symm1}
\text{IR symmetry}&\Rightarrow\text{UV symmetry}\\
\U(1)_{\frac{3V\pm A}{2}}&\Rightarrow \dsZ(\tfrac{3V\pm A}{2}):
\left\{\begin{array}{l}
c_{iA}\to\e^{\pm 3 \ii q_A k_F n}c_{(i+n)A}, \\
c_{iB}\to\e^{\pm 3 \ii q_B k_F n}c_{(i+n)B}.
\end{array}\right.}
Here the {compact} $\U(1)$ symmetries in the IR get mapped to the {non-compact} symmetries $\dsZ$ in the UV, because the UV symmetries are parameterized by the integer variable $n\in\dsZ$. Given that $q_A=1$ and $q_B=3$, \eqnref{eq: 3450 symm1} becomes \eqnref{eq: 3450 symm}, as claimed in the main text. Therefore the 3-4-5-0 chiral fermion model is indeed realized in by the (1+1)D two-band lattice model at low energy.

\section{Wang-Wen Interaction}\label{app: wangwen}

In the bonsonization language, the Wang-Wen interaction is described by
\eq{\scL_\text{int}=\sum_{\alpha=1,2}g_{\alpha}\cos(l_\alpha^\intercal\varphi)}
with $\varphi=(\varphi_{AR},\varphi_{BR},\varphi_{BL},\varphi_{AL})^\intercal$ and the interaction vectors given by
\eq{
l_1=\smat{1\\-2\\1\\2},\quad
l_2=\smat{2\\1\\-2\\1}.}
Mapping back to the chiral fermions by the correspondence $c_{a}\sim{:\,}\e^{\ii\varphi_a}{\,:}$, the interaction reads
\eqs{\label{eq: ww int1}
H_\text{int}&= \frac{g_1}{2}(c_{AR}c_{BL})(c_{BR}^\dagger c_{AL})^2+\text{h.c.}\\
&+\frac{g_2}{2}(c_{BR}c_{AL})(c_{AR} c_{BL}^\dagger)^2+\text{h.c.}.}
According to \eqnref{eq: def ca} and use the inverse Fourier transformation,
\eqs{\label{eq: inv FT}
c_{AR}&=c_{(3k_F)A}=\sum_{i}c_{iA}\e^{3\ii k_F i},\\
c_{BR}&=c_{(-k_F) B}=\sum_{i}c_{iB}\e^{-\ii k_F i},\\
c_{BL}&=c_{(k_F)B}=\sum_{i}c_{iB}\e^{\ii k_F i},\\
c_{AL}&=c_{(-3k_F)A}=\sum_{i}c_{iA}\e^{-3\ii k_F i}.} 
Plugging \eqnref{eq: inv FT} into \eqnref{eq: ww int1}, the interaction becomes
\eq{\label{eq: ww int2}
H_\text{int}=\sum_{i_1,\cdots,i_6}g_{i_1\cdots i_6}(c_{i_1B}^\dagger c_{i_2A})(c_{i_3B}c_{i_4A})(c_{i_5B}^\dagger c_{i_6A})+\text{h.c.},}
with
\eqs{g_{i_1\cdots i_6}&=\frac{g_1}{2}\e^{\ii k_F(i_1-3i_2+i_3+3i_4+i_5-3i_6)}\\
&+\frac{g_2}{2}\e^{\ii k_F(-i_1+3i_2-i_3-3i_4-i_5+3i_6)}.}
Notice that under lattice reflection symmetry $\dsZ_2:c_{iA}\to c_{(-i)A}, c_{iB}\to c_{(-i) B}$, $g_1$ and $g_2$ maps to each other. To simplify, we can impose the reflection symmetry and requires $g_1=g_2=g$, then the coupling coefficient is
\eqs{g_{i_1\cdots i_6}=g\cos\big(k_F(i_1-3i_2+i_3+3i_4+i_5-3i_6)\big).}
The dominant $s$-wave interaction is given by
\eq{\label{eq: icond} i_1-3i_2+i_3+3i_4+i_5-3i_6=0,}
such that $g_{i_1\cdots i_6}=g$ is uniform.
We seek a local interaction that has minimal span on the lattice. The optimal solution of \eqnref{eq: icond} is given by
\eq{\label{eq: isol} i_1=i_2=i-1, i_3=i_4=i, i_5=i_6=i+1,}
for any choice of $i$. With this solution \eqnref{eq: isol}, \eqnref{eq: ww int2} reduces to
\eq{\label{eq: ww int3}
H_\text{int}=g\sum_{i}c_{(i-1)B}^\dagger c_{(i-1)A}c_{iB}c_{iA}c_{(i+1)B}^\dagger c_{(i+1)A}+\text{h.c.}.
}
which is the SMG interaction \eqnref{eq: H two-band interaction} proposed in the main text.

\section{Full Renormalization Group Equations}\label{app:fullrge}

We start with the interaction $\hcf$
\begin{align}
    &\hcf = g_\text{rs} \sum_\alpha \epsilon^{ijk} c_{B \alpha}^\dagger c_{A_i \alpha}c_{A_j \alpha}c_{A_k \alpha} \nonumber \\
    &+g_\text{rt} \sum_{\alpha \ne \beta} \epsilon^{ijk} c_{B \alpha}^\dagger c_{A_i \alpha}c_{A_j \beta}c_{A_k \beta}  +\text{h.c.}
\end{align}
Under RG, the following the density-density and exchange interactions will be generated
\begin{align}
    &\hda =g_\text{as} \sum_{\alpha, s t}  n_{A_{s}\alpha}n_{A_{t}\alpha} +
    g_\text{at} \sum_{\alpha\ne \beta,  s t}  n_{A_{s}\alpha}n_{A_{t}\beta}\nonumber\\
    &+g_\text{ae} \sum_{\alpha\ne \beta,  s t} c_{A_{s}\alpha}^\dagger c_{A_{s}\beta} c_{A_{t} \beta}^\dagger c_{A_{t} \alpha} +(A_{s}\leftrightarrow A_{t})+\text{h.c.}
\end{align}
and
\begin{align}
    &\hdb =g_\text{bs} \sum_{\alpha,s}  n_{B\alpha}n_{A_s\alpha} +
    g_\text{bt} \sum_{\alpha\ne \beta,s}  n_{B\alpha}n_{A_s\beta}\nonumber\\
    &+g_\text{be} \sum_{\alpha\ne \beta,s} c_{B\alpha}^\dagger c_{B\beta} c_{A_s \beta}^\dagger c_{A_s \alpha} +(A_s\leftrightarrow B)+\text{h.c.}
\end{align}
There is an additional density-density interaction that will correct $\hda,\hdb$,
\begin{equation}
    \hbb = g_\text{bb} \sum_{\alpha\beta} n_{B\alpha}n_{B\beta} - c_{B\alpha}^\dagger c_{B\beta} c_{B\beta}^\dagger c_{B\alpha}.
\end{equation}
Putting all interactions together, the full RG equations are

\begin{align*}
 &\frac{dg_{\text{bb}}}{d\ell}=4 d_0 d_{\text{BB}} g_{\text{bb}}^2 +3 d_0 g_{\text{be}}^2 \\
 &\frac{dg_{\text{bs}}}{d\ell}=-2 d_{\text{AB}} g_{\text{bs}}^2+\frac{9 g_{\text{rs}}^2}{2}+g_{\text{rt}}^2 \\
 &\frac{dg_{\text{bt}}}{d\ell}=2 d_0 d_{\text{AB}} g_{\text{bt}}^2 \\
 &\frac{dg_{\text{be}}}{d\ell}=-6 d_0 g_{\text{ae}} g_{\text{be}}+2 d_0 g_{\text{at}} g_{\text{be}}+4 d_0 d_{\text{BB}} g_{\text{bb}}  g_{\text{be}}+3 g_{\text{rs}} g_{\text{rt}}+\frac{g_{\text{rt}}^2}{2} \\
 &\frac{dg_{\text{as}}}{d\ell}=-2 g_{\text{as}}^2 \\
 &\frac{dg_{\text{at}}}{d\ell}=2 d_0 g_{\text{at}}^2-d_0 d_{\text{AB}} g_{\text{rt}}^2 \\
 &\frac{dg_{\text{ae}}}{d\ell}=-d_0 d_{\text{AB}} g_{\text{rt}}^2+4 d_0 g_{\text{ae}} g_{\text{at}}-6 d_0 g_{\text{ae}}^2-2 d_0 d_{\text{BB}} g_{\text{be}}^2 \\
 &\frac{dg_{\text{rs}}}{d\ell}=-6 g_{\text{as}} g_{\text{rs}} \\
 &\frac{dg_{\text{rt}}}{d\ell}=4 d_0 d_{\text{AB}} g_{\text{bt}} g_{\text{rt}}-2 g_{\text{as}} g_{\text{rt}} 
\end{align*}
where $d_{\text{AB}}=d\chi_{pp,\text{AB}}(\vect{0})/d\ell, d_{\text{BB}}=d\chi_{pp,\text{BB}}(\vect{0})/d\ell$. These ratios depend on the energies of $A$ and $B$-type fermions near the VHSs. The two types of fermions have similar band structures, which can be approximated as $E_\vect{k}^{A,B}= \epsilon^{A,B}f(\vect{k})$. The ratios are then given by $d_{\text{AB}}=\frac{2\abs{\epsilon^A}}{\abs{\epsilon^A}+\abs{\epsilon^B}}$ and $d_{\text{BB}}=\frac{\abs{\epsilon^A}}{\abs{\epsilon^B}}$. If $A$ and $B$-type fermions have the same band structure, then $d_{\text{AB}}=d_{\text{BB}}=1$.

\end{document}